# A Machine Learning Method for Prediction of Multipath Channels

Julian Ahrens[1], Lia Ahrens[1], and Hans D. Schotten[1,2]
(1. German Research Center for Artificial Intelligence, 67663 Kaiserslautern, Germany;
2. Technical University of Kaiserslautern, 67663 Kaiserslautern, Germany)

**Abstract**
In this paper, a machine learning method for predicting the evolution of a mobile communication channel based on a specific type of convolutional neural network is developed and evaluated in a simulated multipath transmission scenario. The simulation and channel estimation are designed to replicate real-world scenarios and common measurements supported by reference signals in modern cellular networks. The capability of the predictor meets the requirements that a deployment of the developed method in a radio resource scheduler of a base station poses. Possible applications of the method are discussed.

## 1 Introduction

Today's mobile communication networks are driven by the demand of a steadily increasing number of subscribers for ever higher data rates. This demand has led to the introduction of support for technologies such as millimeter wave transmissions and massive multiple-input multiple-output (MIMO) into the current 5G standards. Apart from the introduction of these new technologies, the available spectrum has to be utilised in the most efficient manner possible. This has already led to the move to orthogonal frequency-division multiple access (OFDMA) and orthogonal frequency-division multiplexing (OFDM) in the fourth-generation mobile broadband standards, which allow for fine grained control over the utilisation of the available radio resources across both the time and frequency domains. While OFDM combats frequency-selective fading by using long symbol times, OFDMA provides further benefits by allowing multiple users to schedule transmission on the subcarriers which are best for them at the time [1]. OFDM also allows for different encodings to be used across the available spectrum, thereby giving the scheduler fine grained control over the trade-off transmission data rate vs. signal robustness.

The dynamic allocation of radio resources and its scheduling is key to achieving efficient utilisation of the available spectrum. Since base stations manage a large number of transmissions, each across a different channel depending on the position and environment of the individual user equipment (UE), they are natural candidates for hosting an optimisation through dynamic scheduling of the radio resources. To achieve efficient radio resource management, scheduling algorithms need to have information about the current and future states of the transmission channels. In particular, two things are required: On the one hand, a mechanism for the estimation of the transmission channels needs to be in place, i.e., there has to be a measurement of the channel transfer function; on the other hand, the development of the transmission channels over time has to be predicted to allow for estimates of future channel quality.

In Long Term Evolution (LTE) systems, channel estimation can be implemented by observing the Cell-Specific Reference Signals (CRS). LTE release 10 (LTE Advanced) supplemented the CRS by the introduction of Channel State Information Reference Signals (CSI-RS). 5G New Radio (NR) does not provide CRS, instead relying exclusively on the flexibly configurable CSI-RS. In this paper, we use a simulation of a multipath propagation transmission channel based on the empirical evidence and the models devised in [2]. The channel is estimated by transmitting and measuring a test signal containing a similar amount of information as the LTE CRS. In particular, very similar estimates can be derived from the observation of LTE CRS.

The present work focusses on the aspect of predicting the time-variant transmission channels. A convolutional neural network (CNN) operating on the time-frequency domain and using multiple time resolutions is designed in order to achieve the necessary prediction performance. The proposed CNN is a two-dimensional variant of the WaveNet network architecture proposed in [3] and uses dilated kernels on the time axis to achieve the incorporation of multiple time resolutions. A further enhancement to the WaveNet architecture presented here consists in enabling simultaneous multi-step predictions, allowing for the instantaneous predictions of the channel development over a period of 5 ms (one half-frame) at a resolution of 500 µs (one slot) each. This is especially useful, since the allocation of resource blocks can be changed at the half-frame level, necessitating the prediction over at least this time period.

The remainder of this work is structured as follows: Section 2 introduces the simulation from which the transfer functions of a fading channel scenario based on real-world observations are derived. Section 3 describes the employed channel estimation procedure. Section 4 describes the channel predictor that is the essential part of this work. Section 5 summarises the obtained results. In Section 6, we provide a discussion of possible applications and an outlook on future research. Section 7 concludes the paper.

## 2 Simulation

Setting the position and carrier frequency of the transmitter to $(0,0) \in \mathbb{R}^2$ and $f_{\text{carrier}} = 900$ MHz, respectively, the multipath transmission is simulated by generating 256 scatterers. Each scatterer starts at a randomly chosen initial position $(x_0^\iota, y_0^\iota) \in \mathbb{R}^2$ such that the power delay profile of the resulting multipath transmission matches the typical urban scenario described in [2], and moves at a random time-invariant velocity $(v_x^\iota, v_y^\iota) \in \mathbb{R}^2$ with $v_x^\iota, v_y^\iota \sim \mathcal{N}(0, \sigma^2)$, $\sigma = 10$ m/s, for $\iota = 0, \dots, 63$, and $v_x^\iota, v_y^\iota = 0$ for $\iota \geq 64$. The receiver is supposed to move from an initial position $(x_0^*, y_0^*) = (400, 0) \in \mathbb{R}^2$ near the transmitter at velocity $v^* = (v_x^*, v_y^*) \in \mathbb{R}^2$ with $|v^*| = 10$ m/s, $\arctan_2(v_y^*, v_x^*) \sim \mathcal{U}(-\pi, \pi)$ where $\exp(i \arctan_2(y, x)) = (x + iy)/\sqrt{x^2 + y^2}$ for $(x, y) \in \mathbb{R}^2$. The transmissions are assumed to be conducted periodically in blocks. The time for transmitting one block is assumed to be $T = 500$ µs, which leads to a discrete time simulation with step size 500 µs. The simulation time amounts to $2^{12} = 4\,096$ time steps in total. The bandwidth of transmission is set to 12.8 MHz. A time interval of length 20 µs at the beginning of each block is used for the transmission of a test signal generated for the channel estimation. All values are computed and stored using International System of Units (SI) base units.

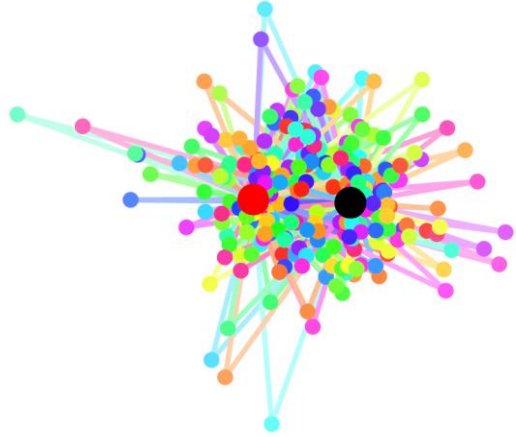

Figure 1. Example configuration of the designed simulation.

An example configuration of this simulation is shown in **Fig. 1**. The large red and black dots represent the transmitter and receiver, respectively. The smaller dots represent the scatterers which are coloured according to the phase offsets observed on the corresponding transmission path (shown as lines) with red representing zero offset and cyan representing a phase offset of $\pi$.

At each simulation step $t$ and for each $\iota$, the path length reflected by the $\iota$-th scatter $l_t^{(\iota)} = |(x_t^\iota, y_t^\iota) - (0,0)| + |(x_t^*, y_t^*) - (x_t^\iota, y_t^\iota)|$ and its derivative with respect to time $\frac{d}{dt} l_t^{(\iota)}$ are computed. For each scatterer $\iota$, we recorded the corresponding transmission delay time $\sigma_t^{(\iota)}$, the constant phase offset $\theta_t^{(\iota)}$, and the Doppler frequency $f_{D_t}^{(\iota)}$ caused by the $\iota$-th scatterer following the rules $\sigma_t^{(\iota)} = l_t^{(\iota)}/c_0$, $\theta_t^{(\iota)} = \left(\left(-l_t^{(\iota)} f_{\text{carrier}}/c_0\right) \bmod 1\right) \cdot 2\pi$, and $f_{D_t}^{(\iota)} = -\frac{d}{dt} l_t^{(\iota)} f_{\text{carrier}}/c_0$, respectively, as well as the received signal amplitude $a_t^{(\iota)}$ computed using the free-space propagation model $a_t^{(\iota)} = c_0/(4\pi f_{\text{carrier}} l_t^{(\iota)})$ ([1]). (Here, $c_0$ refers to the speed of light in vacuum.)

In a setting without line of sight, using linearisation of the phase offset with respect to the Doppler frequency, the time-variant channel impulse response evaluated at time $t + \tau$ for each simulation step $t$ and small $\tau$ resulting from the multipath transmission simulated using the above parameters can be approximated by

$$h(\cdot, t + \tau) = \frac{1}{\sqrt{\sum_{\iota=0}^{255} \left(a_t^{(\iota)}\right)^2}} \sum_{\iota=0}^{255} a_t^{(\iota)} \exp\left(i\theta_t^{(\iota)} + i2\pi f_{D_t}^{(\iota)} \tau\right) \delta_{\sigma_t^{(\iota)}}(\cdot). \tag{1}$$

For any signal $\{S_\tau\}_{0 \leq \tau < T}$ being transmitted in the block beginning at time step $t$ through the simulated channel, this consideration leads to a received signal $\{R_\tau\}_{0 \leq \tau < T}$ in the form of

$$R_\tau = (h(\cdot, t + \tau) * S.)(\tau) = \frac{1}{\sqrt{\sum_{\iota=0}^{255} \left(a_t^{(\iota)}\right)^2}} \sum_{\iota=0}^{255} a_t^{(\iota)} \exp\left(i\theta_t^{(\iota)} + i2\pi f_{D_t}^{(\iota)} \tau\right) \left(\delta_{\sigma_t^{(\iota)}}(\cdot) * S.\right)(\tau). \tag{2}$$

This parametrisation is used in [4] and delivers a realistic approximation of real-world scenarios for numbers of summands greater than 100 [5]. In order to allow continuous time delays to be applied to discrete time signals, the impulse functions $\delta_{\sigma_t^{(l)}}(\cdot)$ in (1) and (2) are convolved with a windowed sinc($\cdot$) function scaled with a given bandwidth. Overall, the channel transmission including pulse shaping with bandwidth restricted to half the sample rate and additive noise is approximated by replacing the $\delta_{\sigma_t^{(l)}}(\cdot)$ in (1) and (2) by $\sin(\pi(\cdot/2))/(\pi(\cdot/2))\mathbf{1}_{[-8,8]}$ and adding independent and identically distributed Gaussian white noise $\sim \mathcal{N}(0, \sigma^2)$ to the transmitted signal with power $\sigma^2$ resulting in a signal-to-noise ratio of 12 dB.

## 3 Channel Estimation

For both the channel estimation and prediction, we will work in the frequency domain. Apart from the obvious usefulness of frequency domain estimation and prediction for OFDM systems, working in the frequency domain allows for a channel estimation scheme of lower computational complexity compared to equalisers operating in the time domain and requiring matrix inversions. Moreover, the frequency domain mode of operation has some benefits on the predictor further detailed in Section 4. Throughout the remainder of this paper, for a discrete time complex-valued signal $\{X_\tau\}_{\tau=0,\dots,N-1}$, let $\mathcal{F}X = \{\mathcal{F}X_f\}_{f=0,\dots,N-1}$ denote its (discrete) Fourier transform.

The time-variant channel transfer functions $\mathcal{F}h(\cdot, t+\tau)$ for $t = 0, \dots, 4\,095T$ and $0 \leq \tau < T$ simulated in Section 2 are approximated by a time series of block wise time-invariant transfer functions $\{\mathcal{F}h^t\}_{t=0,\dots,4\,095}$ based on which the estimation and prediction of the channel transmission are conducted. For each transmission block beginning at time step $t$, in order to estimate the corresponding channel transfer function $\mathcal{F}h^t$, a complex-valued (white noise) test signal $\{\tilde{S}_\tau^t\}_{\tau=0,\dots,N-1}$ whose Fourier transform has constant amplitude and random phases $\sim \mathcal{U}(-\pi, \pi)$ is generated and then transmitted through the multipath transmission channel with additive white noise simulated in Section 2, resulting in a received signal $\{R_\tau^t\}_{\tau=0,\dots,N-1}$. The transfer function $\mathcal{F}h^t$ is in a first step estimated by

$$\mathcal{F}\tilde{h}^t := \frac{\mathcal{F}R^t}{\mathcal{F}\tilde{S}^t}. \qquad (3)$$

In order to improve the quality of the preliminary estimator $\mathcal{F}\tilde{h}^t = \{\mathcal{F}\tilde{h}^t{}_f\}_f$ which is noise corrupted, the correponding impulse response $\tilde{h}^t$ is windowed by a step function of width $N/2$ and then Fourier transformed, i.e., the estimator $\mathcal{F}\hat{h}^t$ of the channel transfer function is given by

$$\mathcal{F}\hat{h}^t = \mathcal{F}\big(\mathbf{1}_{[0,N/2]}\mathcal{F}^{-1}\mathcal{F}\tilde{h}^t\big), \qquad (4)$$

where $\mathcal{F}^{-1}Y = \{\mathcal{F}^{-1}Y_\tau\}_{\tau=0,\dots,N-1}$ refers to the inverse Fourier transform of the considered signal $\{Y_f\}_{f=0,\dots,N-1}$ in the frequency domain. The step of windowing the preliminarily estimated impulse response $\tilde{h}^t = \mathcal{F}^{-1}\mathcal{F}\tilde{h}^t$ is conducted due to the observation of a long noisy tail showing up in the recorded $\tilde{h}^t$, which, according to the simulation with maximum transmission delay time less than $N/2$, should be eliminated; this, at the same time, yields a discrete approximation to convolving the estimated channel transfer function with the kernel $\sin(\pi(\cdot/2))/(\pi(\cdot/2))$ so that down sampling with step size 2 (instead of the original step size 1) in the frequency domain delivers an error corrected version of the estimated channel transfer function $\{\mathcal{F}\hat{h}^t_{2f}\}_{f=0,\dots,N/2-1}$.

When applied to a multipath transmission channel with additive white noise such as the transmission channel simulated in Section 2, the above method of estimating the channel transfer function yields a reasonably accurate estimate.

The initial resolution level of the frequency spectrum is set to $N = 2^9$ which results in an estimated channel transfer function $\mathcal{F}\hat{h}^t$ of length $N/2 = 256$ for each block beginning at simulation step $t$. Overall, the simulation is run 16 times independently, which results in 16 independent time series of the form $\{\mathcal{F}\hat{h}^t\}_{t=0,\dots,4\,095}$ with $\mathcal{F}\hat{h}^t \in \mathbb{C}^{256}$.

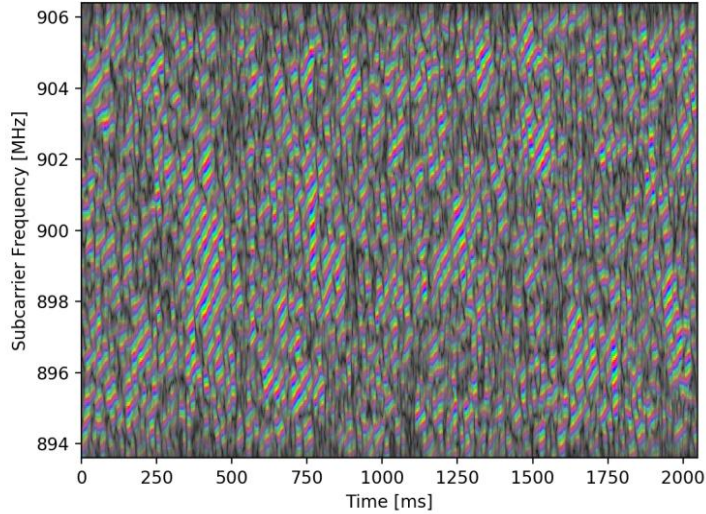

Figure 2. Example plot of estimated transfer functions.

An example realisation of transfer functions estimated during one simulation is shown in Fig. 2. The $x$-axis represents time, labelled by time steps $t$, and the $y$-axis represents frequency, labelled by indices of subcarriers $f$. Brightness corresponds to amplitude with bright colors representing good signal reception. Colors correspond to phases with red representing phase 0 and cyan representing phase π. One can clearly see the thin dark areas reflecting the effect of frequency selective fading.

In order to ensure that the proposed system could indeed be implemented on current cellular radio equipment, the method of estimation of the simulated channel is chosen in such a manner that the level of channel information obtained is very similar to that commonly available from the reference signals in real-world systems.

## 4 Channel Prediction

The time series of estimated channel transfer functions from Section 3 are used as labels for training and testing a carefully chosen convolutional neural network (CNN) that delivers one- or multi-step ahead predictions of the time-variant channel transfer functions resulting from the simulation in Section 2. Since additive noise is included in the simulation of the channel, the trained neural network also contributes to the denoising of the channel transfer function along with the channel estimation scheme in Section 3.

In general, CNNs are a specific architecture of feed-forward neural networks, where linear filters (convolution kernels) instead of traditional single weight parameters are used in a shift invariant manner for the transformation between adjacent layers, making use of local temporal and spatial structure of the input signal within a local receptive field. The local receptive field can be enlarged without the need for increasing the number of parameters by means of the dilation parameter. In the one-dimensional case, a CNN with dilation is known as WaveNet that is introduced in [3] for processing audio signals. For more details on CNNs, the readers are referred to [6]. Compared to traditional fully connected neural networks and recurrent neural networks such as long short-term memory units (LSTMs) [7], CNNs use fewer parameters and are less receptive to overfitting.

The shift invariant nature of CNNs necessitates that the signals processed by a CNN have some amount of homogeneity, as the layers of the CNN have no way of varying the processing performed by them between different regions of the input signal. In our particular case, this means that the method in which predictions are performed for a certain consecutive group of subcarriers is exactly the same as that used for any other group of consecutive subcarriers. This is a reasonable approach, as the manner in which the influence of the channel on the transmission develops over time is indeed very homogeneous across the entire considered bandwidth. This assumption would not hold, if we were to work directly on the time domain channel impulse response, as most of the power of this impulse response is contained within the first few microseconds, suggesting a different approach for processing this earlier part of the impulse response.

In our setting, a two-dimensional convolutional neural network (CNN) with partial dilation is used for building the prediction model, which is described in the remainder of this section.

CNNs are a special type of feed-forward neural networks made up of one or several convolutional layers. A feed-forward neural network is a function mapping an input vector to an output vector, making use of a set of parameters which are to be adapted through the training. In a multi-layer neural network, this function operates in the form of several such functions in succession, each transforming the corresponding input vector into an output vector. In our setting, for processing the time series of channel transfer functions $\{\mathcal{F}\hat{h}^t\}_{t=0,1,...}$ with $\mathcal{F}\hat{h}^t \in \mathbb{C}^{256}$, we use two-dimensional convolutional layers where each input vector is indexed with three axes related to the real-or-imaginary part of the complex plane, the simulation time steps, and the frequency domain, and the transformation is conducted by convolving the input vector with a convolution kernel made up of free parameters to be adapted and adding a free parameter vector called bias to the result. For our purpose of multi-step prediction, we also consider the evolution of the time series over a long period of time, for which we use the so-called dilation parameter on the time axis defining the spacing between the free parameters in the convolution kernel. The introduction of the dilation parameter enables us to extend the receptive field of the CNN in time without taking extra parameters for fitting.

For delivering at most $m$-step ahead predictions of the future channel transfer function, we use a 5-layer CNN beginning with 4 consecutive partially dilated convolutional layers along the $t$-axis with channel sizes 2, 6, 12, 12, 6, followed by one convolutional layer with $2m$ output channels. In each partially dilated convolutional layer, the size of the free convolutional kernel is set to $(4, 5)$ for the time and frequency axis, respectively, and the dilation parameter is defined by 4 to the power of the corresponding layer number. The final layer is endowed with $1 \times 1$ convolution kernels. Apart from the last layer, the hyperbolic tangent is used as activation function in each layer. In order to improve the back propagation of the gradient ([8]), a residual convolutional layer ([9]) with kernel size $(1, 1)$ is added to each partially dilated convolutional layer. The above layout is common in convolutional neural networks and is designed to best adapt to our task and the nature of the input signals. The layout of our CNN is summarised in Table 1. For illustration, a diagram of the dilated layers along the $t$-axis is presented in Fig. 3.

Table 1. Layout of the 2D CNN for $m$-step ahead prediction

| Layer number | Layer type | Channel sizes | Kernel size $(t, f)$ | Dilation parameter | Residual layer |
|---|---|---|---|---|---|
| 0 | dilated convolution | $2 \rightarrow 6$ | $(4, 5)$ | $(1, 1)$ | $1 \times 1$ convolution |
| 1 | dilated convolution | $6 \rightarrow 12$ | $(4, 5)$ | $(4, 1)$ | $1 \times 1$ convolution |
| 2 | dilated convolution | $12 \rightarrow 12$ | $(4, 5)$ | $(16, 1)$ | $1 \times 1$ convolution |
| 3 | dilated convolution | $12 \rightarrow 6$ | $(4, 5)$ | $(64, 1)$ | $1 \times 1$ convolution |
| 4 | convolution | $6 \rightarrow 2m$ | $(1, 1)$ | $(1, 1)$ | None |

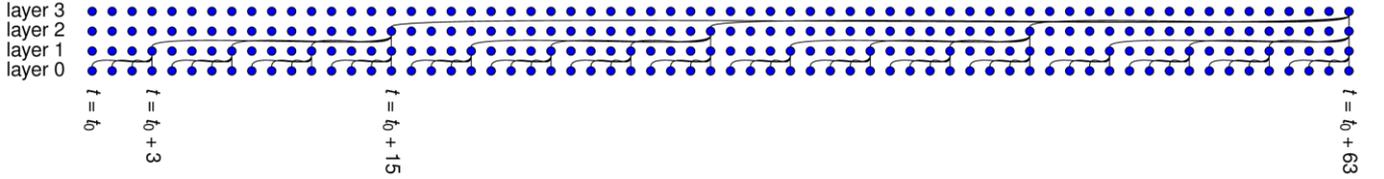

Figure 3. Structure of the dilated convolutional layers along the $t$-axis.

During the training, the free parameters in our CNN are adjusted to the labelled training data by minimising the mean squared error (MSE) of prediction along the negative direction of the gradient of the error function with respect to the parameters, for which we use a refined version of stochastic gradient descent (SGD) called ADAM training algorithm ([10]). The gradient for each update is computed by means of the so-called backpropagation algorithm ([8]) based on the chain rule.

In our setting, the 16 independent time series of channel transfer functions $\{\mathcal{F}\hat{h}^t\}_{t=0,\ldots,4\,095}$ with $\mathcal{F}\hat{h}^t \in \mathbb{C}^{256}$ are each divided into 8 segments which are to be fed into the CNN as input vectors and grouped as training, validation, and test parts with the proportion $6 : 1 : 1$. The ADAM optimiser with learning rate $\gamma = 0.01$ is run for 30 training epochs in total.

## 5 Results

The performance of our approach to delivering multi-step ahead prediction is measured in a setting with $m = 10$ for training the corresponding CNN to output 10-step ahead predictions at most. The MSEs are evaluated for training, validation, and test data (Table 2). The similarity of performance evaluated on all three sub-datasets indicates no significant overfitting.

Table 2. Mean squared errors (MSEs) for prediction length $\Delta t$ from 1 to 10

| $\Delta t$ | Training | Validation | Test | Trivial |
|---|---|---|---|---|
| 1 | 0.1466 | 0.1460 | 0.1424 | 0.2217 |
| 2 | 0.1579 | 0.1569 | 0.1538 | 0.2532 |
| 3 | 0.1753 | 0.1735 | 0.1707 | 0.3047 |
| 4 | 0.2014 | 0.1988 | 0.1962 | 0.3750 |
| 5 | 0.2357 | 0.2321 | 0.2298 | 0.4621 |
| 6 | 0.2757 | 0.2723 | 0.2704 | 0.5644 |
| 7 | 0.3249 | 0.3208 | 0.3183 | 0.6788 |
| 8 | 0.3821 | 0.3763 | 0.3735 | 0.8040 |
| 9 | 0.4459 | 0.4393 | 0.4354 | 0.9382 |
| 10 | 0.5133 | 0.5055 | 0.5005 | 1.0797 |

As a baseline, we consider the trivial prediction where all future values of the time series are set to the latest observed value; the MSE of such a prediction scheme provides a measure for the variation of the underlying time series over time (Table 2). Overall, the instantaneous long-term prediction with our CNN using $m = 10$ facilitated by employing the dilation parameter in time delivers much more accurate results than the trivial prediction scheme.

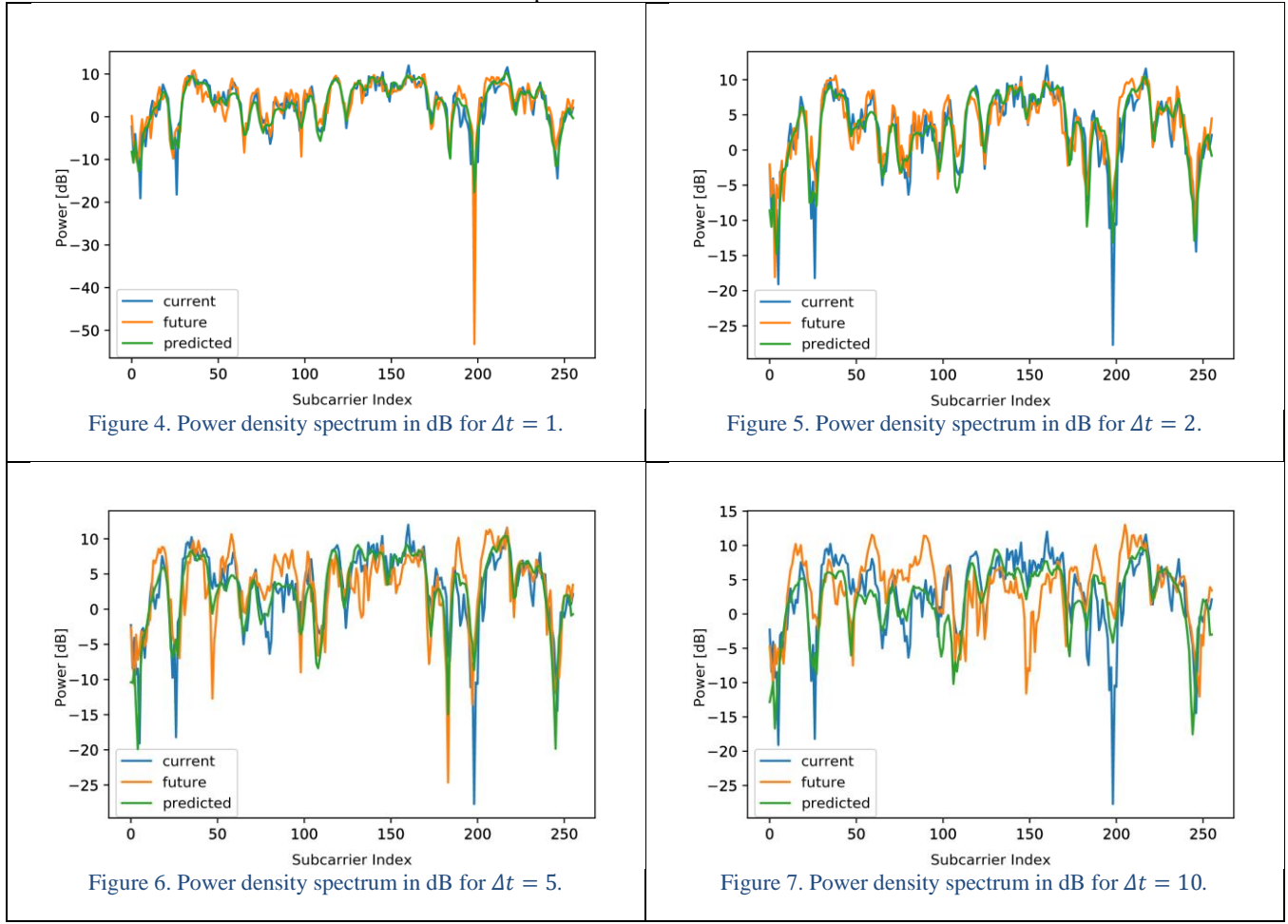

Figure 4. Power density spectrum in dB for $\Delta t = 1$.

Figure 5. Power density spectrum in dB for $\Delta t = 2$.

Figure 6. Power density spectrum in dB for $\Delta t = 5$.

Figure 7. Power density spectrum in dB for $\Delta t = 10$.

In Figs. 4, 5, 6, and 7, the power density spectra in dB of an example channel transfer function evaluated at time steps $t_0$ and $t_0 + \Delta t$ and the $\Delta t$-step ahead instantaneous prediction for time $t_0 + \Delta t$, $\Delta t = 1, 2, 5, 10$, output by the trained CNN with $m = 10$, are plotted in blue, yellow, and green, respectively. Note in particular that most of the negative peaks of the future power density spectra are correctly detected by the predictor, which suggests the utility of our approach in handling situations with frequency selective fading in an OFDM transmission scheme (see Section 6 for more discussion).

## 6 Discussion

As mentioned in the introduction, the method proposed in the preceding sections can be employed to provide an OFDMA/OFDM radio resource scheduler located in a base station with predictions necessary for an efficient scheduling of radio resources. There are two main aspects of the scheduler, which can benefit from this information:

The predictions can be used to decide to which user a specific radio resource element should be allocated by estimating the relative usefulness of assigning the element to a specific user compared to the utility another user may have of it. For instance, consider the case where two radio resource blocks RRB A and RRB B are assigned to users UE A and UE B, respectively. If the predictor predicts that during the next half-frame the part of the spectrum on which RRB A is transmitted will become faded for UE A, but a strong signal could be received by UE B, it would be advantageous to change the allocation and assign RRB B to UE A and RRB A to UE B.

The other aspect is that the scheduler may control the choice of encoding used on each of the radio resource elements. In particular, if a prediction reveals that a certain part of the spectrum will become faded for a particular user and a reallocation among the users as in the case discussed previously is not applicable, the scheduler may initiate a change of the employed encoding, for instance from 64QAM[1] down to 16QAM, thereby increasing the robustness of the signal and counteracting the decreasing signal-to-noise ratio. In extreme cases of frequency selective fading, transmissions on the corresponding frequencies could even be disabled completely.

In future research, we hope to expand on both of these topics by developing an adaptive coding scheme and a dynamic scheduler for the multi-user case based on the research performed in this article.

---

[1] Quadrature amplitude modulation

# 7 Conclusions

In this paper, we simulated a multipath transmission scenario, implemented a channel estimation scheme, and designed a machine learning model for predicting the resulting channel transfer functions over multiple time steps. Our results show that the machine learning model is capable of capturing characteristics of the channel evolution and provides reasonable predictions. We addressed possible applications of the method in real-world systems, which we plan to implement and evaluate in future research.

# Biographies

**Julian Ahrens** (Julian.Ahrens@dfki.de) received his Master's degree in mathematics from Kiel University (CAU), Germany, while working in non-commutative harmonic analysis. He is currently working as a researcher at the Intelligent Networks Group of Prof. Hans D. Schotten at the German Research Center for Artificial Intelligence, where he is involved in the BMBF project Future Industrial Network Architecture (FIND). His research interests include high performance computing, artificial intelligence, digital signal processing, and harmonic and functional analysis.

**Lia Ahrens** received her Ph.D. in Stochastics and Financial Mathematics from Kiel University (CAU), Germany. She is currently working as a senior researcher at the Intelligent Networks Group of Prof. Hans D. Schotten at the German Research Center for Artificial Intelligence, where she is involved in the BMBF project TACNET 4.0—Taktiles Internet. Her research interests include stochastic processes, stochastic filtering, and machine learning for time series analysis.

**Hans D. Schotten** received his Diploma and Ph.D. degrees in electrical engineering from the RWTH Aachen University of Technology, Germany. He is a full professor and director of the Institute for Wireless Communications and Navigation at the Technical University of Kaiserslautern, Germany. In addition, he is a scientific director of the German Research Center for Artificial Intelligence (DFKI) and head of the department for Intelligent Networks. Since 2018, he is the chairman of the German Society for Information Technology and member of the Supervisory Board of the VDE.